\documentclass[prd,twocolumn,showpacs,preprintnumbers,superscriptaddress,floatfix,nofootinbib]{revtex4-1}
\usepackage{amsmath,amssymb}
\usepackage{graphicx,bm,braket}
\usepackage{subfigure}
\usepackage{color}
\pdfpagewidth=\paperwidth
\pdfpageheight=\paperheight
\allowdisplaybreaks
\newcommand*{\slashed}[1]{{#1\!\!\!/}}
\newcommand*{\hc}{\text{H.\,c.}}

\begin{document}

\title{\boldmath Analysis of the data for $\gamma p \to f_1(1285) p$ photoproduction}

\author{Ai-Chao Wang}
\affiliation{School of Nuclear Science and Technology, University of Chinese Academy of Sciences, Beijing 101408, China}

\author{Neng-Chang Wei}
\affiliation{School of Nuclear Science and Technology, University of Chinese Academy of Sciences, Beijing 101408, China}

\author{Fei Huang}
\email[Corresponding author: ]{huangfei@ucas.ac.cn}
\affiliation{School of Nuclear Science and Technology, University of Chinese Academy of Sciences, Beijing 101408, China}

\date{\today}

\begin{abstract}
The photoproduction of $f_1(1285)$ meson off proton is investigated within an effective Lagrangian approach. The $t$-channel $\rho$- and $\omega$-exchange diagrams, $u$-channel nucleon-exchange diagram, generalized contact term, and $s$-channel pole diagrams of nucleon and a minimal number of nucleon resonances are taken into account in constructing the reaction amplitudes to describe the experimental data. Three different models, i.e., the Feynman model, the Regge model, and the interpolated Regge model, are employed where the $t$-channel reaction amplitudes are constructed in Feynman type, Regge type, and interpolated Regge type, respectively. The results show that in neither Feynman model with two nucleon resonances nor interpolated Regge model with one nucleon resonance can the available data for $\gamma p \to f_1(1285) p$ be satisfactorily reproduced. Nevertheless, in the Regge model, when any one of the $N(1990){7/2}^+$, $N(2000){5/2}^+$, $N(2040){3/2}^+$, $N(2060){5/2}^-$, $N(2100){1/2}^+$, $N(2120){3/2}^-$, $N(2190){7/2}^-$, $N(2300){1/2}^+$, and $N(2570){5/2}^-$ resonances is considered, the data can be well described. The resulted resonance parameters are consistent with those advocated in Particle Data Group (PDG) review. Further analysis shows that in high-energy region, the peaks of $\gamma p \to f_1(1285) p$ differential cross sections at forward angles are dominated by the contributions from $t$-channel $\rho$- and $\omega$-exchange diagrams, while in low-energy region, the $s$-channel pole diagrams of resonances also provide significant contributions to the $\gamma p \to f_1(1285) p$ cross sections.
\end{abstract}


\maketitle

\section{Introduction}   \label{Sec:intro}

The study of nucleon resonances ($N^\ast$'s) and $\Delta$ resonances ($\Delta^\ast$'s) has always been of great interest in hadron physics community, since a thorough understanding of the structure and properties of $N$ and $\Delta$ resonances is essential to understand clearly the nonperturbative behavior of quantum chromodynamics (QCD), the fundamental theory of strong interactions. It is well known nowadays that the experimental and theoretical studies of $\pi N$ scattering and $\pi$ photoproduction reactions provide us the most knowledge about the $N^\ast$'s and $\Delta^\ast$'s. Nevertheless, quark models \cite{Isgur:1978,Capstick:1986,Loring:2001} predicated much more $N^\ast$'s and $\Delta^\ast$'s than experimentally observed. The possible reason might be that some of the unobserved $N^\ast$'s or $\Delta^\ast$'s couple weakly to $\pi N$ channel and thus escaped from the observation. On the other hand, there are many one-star and two-star $N^\ast$'s and $\Delta^\ast$'s whose parameters especially the decay branching ratios to various final states are not yet known in the most recent Particle Data Group (PDG) review \cite{Workman:2022}. These situations force us to investigate the $N^\ast$'s and $\Delta^\ast$'s in reaction channels other than $\pi$ hadro- and photoproductions. In the past few years, lots of experimental and theoretical efforts have been devoted to the study of $\eta N$, $\eta' N$, $KY$, $\omega N$, $K^\ast Y$, and $KY^\ast$ $(Y=\Lambda, \Sigma)$ photoproduction reactions, and valuable achievements have been obtained \cite{Kashevarov:2017,Anisovich:2018,Zhangy:2021,Bradford:2006,Mart:2019,CLAS-beam,Zachariou:2020kkb,Wei:2022,Wang:2022,Anisovich:2017rpe,Wei:2020,Moriya:2013,Wangx:2020,Zhangyx:2021,Wein:2021,Roy:2018,Weinc:2019}. In the present work, we concentrate on the photoproduction reaction of $f_1(1285)$ meson off the proton target. The $f_1(1285)$ is an axial-vector meson with quantum numbers $I^G J^{PC}=0^+ 1^{++}$, mass $M=1281.9\pm 0.5$ MeV and width $\Gamma=22.7\pm 1.1$ MeV \cite{Workman:2022}. As the $f_1(1285) p$ threshold is above $2.2$ GeV, this reaction is more suitable than $\pi$ production reactions to investigate the $N^\ast$'s with higher mass in the less-explored energy region. Furthermore, the $f_1(1285) p$ photoproduction reaction acts as an ``isospin filter'', isolating the nucleon resonances with isospin $I=1/2$ and eliminating the interference of the $\Delta$ resonances which have isospin $I=3/2$.

Experimentally, the differential cross-section data for $\gamma p \to f_1(1285) p$ with $f_1(1285) \to \eta \pi^+\pi^-$ in the energy range from the $f_1(1285) p$ threshold up to the center-of-mass energy $W\approx  2.8$ GeV were released in $2016$ by the CLAS Collaboration at the Thomas Jefferson National Accelerator Facility (JLab) \cite{Dickson:2016gwc}. Compared with the simultaneously released cross-section data for $\gamma p \to \eta'(958)p$ with $\eta'(958)\to \eta\pi^{+}\pi^{-}$, the $f_1(1285)$ cross sections are much flatter in $\cos\theta$ angular dependence, indicating less important $t$- and $u$-channel contributions and possible $s$-channel nucleon resonance contributions.

Theoretically, several works have been committed to the study of $\gamma p \to f_1(1285) p$ photoproduction reaction. Before the publication of the CLAS data, predictions of the differential cross sections for $\gamma p \to f_1(1285) p$ from various theoretical models were available in Refs.~\cite{Kochelev:2009,Domokos:2009,HuangY:2014}. Nevertheless, none of them seemed to be able to even qualitatively describe the CLAS data reported in 2016 \cite{Dickson:2016gwc}. Since the CLAS differential cross-section data for $\gamma p \to f_1(1285) p$ became available, there were two theoretical works analyzing them, both in the framework of the so-called interpolated Regge model \cite{Wangxiaoyun:2017,Wangyanyan:2017}. However, the conclusions drawn from these two analyses are quite different. In Ref.~\cite{Wangxiaoyun:2017}, all contributions from resonance-pole diagrams, $NN\gamma$ vector coupling ($\propto {\bar N}\gamma_\mu A^\mu N$) in $s$-channel $N$-pole diagram and $u$-channel $N$-exchange diagram, and contact term were omitted, and only the contributions from $t$-channel $\rho$- and $\omega$-exchange diagrams together with those from the $NN\gamma$ tensor coupling ($\propto {\bar N}\sigma^{\mu\nu}\partial_\nu A_\mu N$) in $s$-channel $N$-pole diagram and $u$-channel $N$-exchange diagram were considered in the calculation. It was concluded that the CLAS differential cross-section data can be reproduced without considering any nucleon resonances. In Ref.~\cite{Wangyanyan:2017}, besides the $t$-channel $\rho$- and $\omega$-exchange diagrams, $s$-channel $N$-pole diagram, $u$-channel $N$-exchange diagram, and contact term contributions, the $N(2300){1/2}^+$-pole diagram was further considered in constructing the reaction amplitudes. It was reported that the contribution from the $N(2300){1/2}^+$-pole diagram plays an important role in reproducing the CLAS differential cross-section data.

As both Refs.~\cite{Wangxiaoyun:2017} and \cite{Wangyanyan:2017} used the interpolated Regge model and they analyzed the same set of data, one is confused whether the differences of their conclusions about the $\gamma p \to f_1(1285) p$ reaction mechanisms and the extracted resonance contents are due to the fact that the contributions from contact term and $NN\gamma$ vector coupling ($\propto {\bar N}\gamma_\mu A^\mu N$) were omitted in Ref.~\cite{Wangxiaoyun:2017}. Furthermore, one is curious whether the $N(2300){1/2}^+$ resonance is the only one that is needed to describe the CLAS differential cross-section data for $\gamma p \to f_1(1285) p$ in an interpolated Regge model as done in Ref.~\cite{Wangyanyan:2017}. Besides, one wants to make clear whether the traditional Feynman model and/or Regge model are capable of describing the available cross-section data from the CLAS Collaboration for the $\gamma p \to f_1(1285) p$ reaction.

To eliminate the above-mentioned confusions, in the present work, we present an independent and comprehensive analysis of the CLAS data \cite{Dickson:2016gwc} on differential cross sections for $f_1(1285)p$ photoproduction in an effective Lagrangian approach. We build three different reaction models, i.e., the Feynman model, the Regge model, and the interpolated Regge model, where the reaction amplitudes of $t$-channel $\rho$ and $\omega$ exchanges are constructed in Feynman type, Regge type, and interpolated Regge type, respectively. In each of these three models, the $t$-channel $\rho$- and $\omega$-exchange diagrams, the $s$-channel $N$-pole diagram, the $u$-channel $N$-exchange diagram, and the generalized interaction current are considered as background ingredients, and besides, the $s$-channel pole diagrams of a minimum number of nucleon resonances are considered in order to get a satisfactory description of the available data. The gauge invariance of the photoproduction amplitude is guaranteed by introducing an auxiliary current which ensures that the full reaction amplitude satisfies the generalized Ward-Takahashi identity and thus is fully gauge invariant, independent of any specific type of form factors introduced in hadronic vertices \cite{Haberzettl:1997,Haberzettl:2006,Huang:2012,Huang:2013}.

The present paper is organized as follows. In Sec.~\ref{Sec:formalism}, we introduce the framework of our theoretical model. In Sec.~\ref{Sec:results}, we present our theoretical results and a discussion of them. Finally, a brief summary and conclusions are given in Sec.~\ref{sec:summary}.

\section{Formalism}  \label{Sec:formalism}

\begin{figure}[tbp]
\subfigure[~$s$ channel]{
\includegraphics[width=0.45\columnwidth]{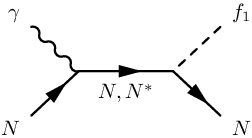}}  {\hglue 0.4cm}
\subfigure[~$t$ channel]{
\includegraphics[width=0.45\columnwidth]{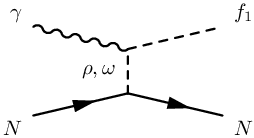}} \\[6pt]
\subfigure[~$u$ channel]{
\includegraphics[width=0.45\columnwidth]{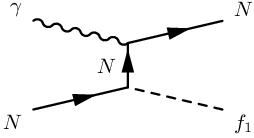}} {\hglue 0.4cm}
\subfigure[~Interaction current]{
\includegraphics[width=0.45\columnwidth]{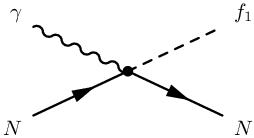}}
\caption{Generic structure of the amplitude for $\gamma N\to f_1(1285) N$. Time proceeds from left to right. The symbols $f_1$ denote $f_1(1285)$.}
\label{FIG:feymans}
\end{figure}

The full photoproduction amplitudes for $\gamma N \to f_1(1285) N$ can be expressed as
\begin{equation}
M^{\nu\mu} = M^{\nu\mu}_s + M^{\nu\mu}_t + M^{\nu\mu}_u + M^{\nu\mu}_{\rm int},  \label{eq:amplitude}
\end{equation}
with $\nu$ and $\mu$ being Lorentz indices of the $f_1(1285)$ meson and the photon, respectively. The first three terms $M^{\nu\mu}_s$, $M^{\nu\mu}_t$, and $M^{\nu\mu}_u$ stand for the $s$-, $t$-, and $u$-channel amplitudes, respectively, with $s$, $t$, and $u$ being the Mandelstam variables of the internally exchanged particles. They arise from the photon attaching to the external particles in the underlying $NNf_{1}$ interaction vertex. The last term, $M^{\nu\mu}_{\rm int}$, stands for the interaction current that arises from the photon attaching to the internal structure of the $NNf_{1}$ interaction vertex. All four terms in Eq.~(\ref{eq:amplitude}) are diagrammatically depicted in Fig.~\ref{FIG:feymans}.

As shown in Fig.~\ref{FIG:feymans}, in the present work, contributions from the following interaction diagrams are considered in constructing the $s$-, $t$-, and $u$-channel reaction amplitudes: (i) $s$-channel $N$- and $N^\ast$-pole diagrams, (ii) $t$-channel $\rho$- and $\omega$-exchange diagrams, and (iii) $u$-channel $N$-exchange diagram. Note that the $u$-channel interaction is expected to contribute significantly at high energy backward angles. However, the available cross section data for $\gamma p\to f_1(1285) p$ are sparse in this energy and angle regime. In this situation, we simply ignore the $u$-channel resonance exchanges in the present work, as the additional parameters introduced with these interactions cannot be well determined by the available data. The expressions for all the $s$-, $t$-, and $u$-channel amplitudes can be obtained straightforwardly by evaluating the corresponding Feynman diagrams, except that in Regge model and interpolated Regge model appropriate substitutions are made for the propagators of intermediate $\rho$ and $\omega$ mesons in the $t$-channel Feynman amplitudes (cf. Sec.~\ref{Sec:Regge_treatments}). The interaction current $M^{\nu\mu}_{\rm int}$ in Eq.~\eqref{eq:amplitude} cannot be calculated directly, and we follow Refs.~\cite{Haberzettl:1997,Haberzettl:2006,Huang:2012,Huang:2013} to model this term by a generalized contact current,
\begin{equation}
M^{\nu\mu}_{\rm int} = \Gamma^\nu_{NNf_{1}}(q) \, C^\mu.  \label{eq:Mint}
\end{equation}
Here $\Gamma^\nu_{NNf_{1}}(q)$ is the vertex function of $NNf_{1}$ coupling given by the Lagrangian of Eq.~(\ref{eq:L_NNf1}),
\begin{equation}
\Gamma^\nu_{NNf_{1}}(q) = i g_{NNf_1} \left(1 + \frac{\kappa_{f_1}}{2m_{N}} \slashed{q} \right) \gamma^\nu \gamma^5,
\end{equation}
with $q$ being the four-momentum of the outgoing $f_{1}$ meson; $C^\mu$ is an auxiliary current, which is nonsingular and is introduced to ensure that the full photoproduction amplitudes of Eq.~(\ref{eq:amplitude}) satisfy the generalized Ward-Takahashi identity and thus are fully gauge invariant. Following Refs.~\cite{Haberzettl:2006,Huang:2012}, we choose $C^\mu$ for $\gamma p \to f_1(1285) p$ as
\begin{equation}   \label{eq:Cmu}
C^\mu = - Q_{N_{u}}\frac{f_{u}-\hat{F}}{u-p'^{2}}(2p'-k)^{\mu}- Q_{N_{s}}\frac{f_{s}-\hat{F}}{s-p^2}(2p+k)^{\mu},
\end{equation}
with
\begin{equation} \label{eq:Fhat-Kstp}
\hat{F} = 1 - \hat{h} \left(1 -  f_u\right) \left(1 - f_s\right).
\end{equation}
Here $p$, $p'$, and $k$ are the four-momenta for the incoming $N$, outgoing $N$, and incoming photon, respectively; $Q_{N_u}$ and $Q_{N_s}$ are electric charges of $u$-channel $N$ and $s$-channel $N$, respectively; $f_u$ and $f_s$ are the phenomenological form factors attached to the amplitudes of $u$-channel $N$-exchange diagram and $s$-channel $N$-pole diagram, respectively; $\hat{h}$ is an arbitrary function that goes to unity in the high-energy limit to prevent the ``violation of scaling behavior'' ~\cite{Drell:1972}. In the present work, we choose $\hat{h}=1$ for the sake of simplicity.

\subsection{Effective Lagrangians} \label{Sec:Lagrangians}

The effective interaction Lagrangians used in the present work for constructing the production amplitudes are given below. For further convenience, we define the operators
\begin{equation}
\Gamma^{(+)}=\gamma_5  \qquad  \text{and} \qquad  \Gamma^{(-)}=1,
\end{equation}
and the field-strength tensors
\begin{eqnarray}
f_1^{\mu\nu} &=& \partial^\mu f_1^\nu - \partial^\nu f_1^\mu,  \\[6pt]
F^{\mu\nu} &=& \partial^\mu A^\nu - \partial^\nu A^\mu,
\end{eqnarray}
with $f_1^\nu$ and $A^\mu$ denoting the $f_1$ vector-meson field and electromagnetic field, respectively.

The electromagnetic interaction Lagrangians for $s$-channel $N$-pole diagram, $u$-channel $N$-exchange diagram, and $t$-channel $\rho$- and $\omega$-exchange diagrams read
\begin{align}
{\cal L}_{NN\gamma} &= -e \bar{N} \left[ \left( \hat{e} \gamma^\mu - \frac{ \hat{\kappa}_N} {2M_N}\sigma^{\mu \nu}\partial_\nu\right) A_\mu\right] N,  \label{eq:L_NNg}    \\[6pt]
{\cal L}_{Vf_1\gamma } &= i g_{Vf_1\gamma} \varepsilon_{ \mu \nu \alpha \beta} \left(\partial ^{\mu } A^{\alpha }\right) \left(\partial^2 V^\nu\right) f_1^\beta,
\end{align}
where $e$ is the elementary charge unit and $\hat{e}$ stands for the charge operator; $\hat{\kappa}_N = \kappa_p\left(1+\tau_3\right)/2 + \kappa_n\left(1-\tau_3\right)/2$, with the anomalous magnetic moments $\kappa_p=1.793$ and $\kappa_n=-1.913$; $M_N$ stands for the masse of $N$; $\varepsilon_{ \mu \nu \alpha \beta}$ is the totally antisymmetric Levi-Civita tensor with $\varepsilon_{0123}=1$; $V$ represents the vector meson $\rho$ or $\omega$. The coupling constant $g_{\rho f_{1}\gamma}$ can be determined by the decay width of $f_1\to\rho\gamma$,
\begin{equation}
\Gamma _{f_1 \to \rho \gamma } =  \frac{ g_{\rho f_1 \gamma }^2 m_{\rho }^{2} }{96\pi m_{f_1}^5} \left(m_{f_1}^2 + m_{\rho }^2 \right) \left(m_{f_1}^2 - m_{\rho }^2 \right)^3.
\end{equation}
With the value $\Gamma _{f_1 \to \rho \gamma } \simeq 0.45$ MeV from the CLAS experiment \cite{Dickson:2016gwc}, one gets $g_{\rho f_1\gamma} = 0.94$ GeV$^{-2}$. The coupling constant $g_{\omega f_1\gamma }$ can be estimated within a quark model assuming that $f_1$ consists of two-flavor $u,d$ quarks, which results in $g_{\omega f_1\gamma } \approx g_{\rho f_1\gamma}/3$ \cite{Kochelev:2009}.

The hadronic interaction Lagrangians for $s$-channel $N$-pole diagram, $u$-channel $N$-exchange diagram, and $t$-channel $\rho$- and $\omega$-exchange diagrams read
\begin{align}
{\cal L}_{NNf_1} &= g_{NNf_1} \bar{N} \left( f_1^\mu - i\frac{\kappa_{f_1}}{2m_N} \gamma^\nu \partial_\nu f_1^\mu \right) \gamma_\mu \gamma^5 N,  \label{eq:L_NNf1}  \\[6pt]
{\cal L}_{NNV} &= -g_{NNV} \bar{N} \left[ \gamma_\mu V^\mu - \frac{\kappa_V}{2m_{N}} \sigma_{\mu\nu}\partial^\nu V^\mu \right] N.
\end{align}
The value of coupling constant $g_{NNf_1}= 2.5$ is obtained according to the relation between the reduced elements $F$ and $D$ from a study of the axial vector currents in Ref.~\cite{Birkel:1996}. The coupling $\kappa _{f_1}$ is treated as a parameter to be fixed by fitting the CLAS experimental data \cite{Dickson:2016gwc}. For nucleon and vector meson couplings, the empirical values $g_{NN\rho }=3.25$, $g_{ NN\omega}=11.76$, $\kappa_{\rho}=6.1$ and $\kappa_{\omega} = 0$ from Refs.~\cite{Huang:2012,Ronchen:2013} are quoted.

The Lagrangians for resonance-nucleon-photon transition read
\begin{eqnarray}
{\cal L}_{RN\gamma}^{1/2\pm} &=& e\frac{g_{RN\gamma}^{(1)}}{2M_N}\bar{R} \Gamma^{(\mp)}\sigma_{\mu\nu} \left(\partial^\nu A^\mu \right) N  + \hc, \\[6pt]
{\cal L}_{RN\gamma}^{3/2\pm} &=& -\, ie\frac{g_{RN\gamma}^{(1)}}{2M_N}\bar{R}_\mu \gamma_\nu \Gamma^{(\pm)}F^{\mu\nu}N \nonumber \\
&&+\, e\frac{g_{RN\gamma}^{(2)}}{\left(2M_N\right)^2}\bar{R}_\mu \Gamma^{(\pm)}F^{\mu \nu}\partial_\nu N + \hc, \\[6pt]
{\cal L}_{RN\gamma}^{5/2\pm} & = & e\frac{g_{RN\gamma}^{(1)}}{\left(2M_N\right)^2}\bar{R}_{\mu \alpha}\gamma_\nu \Gamma^{(\mp)}\left(\partial^{\alpha} F^{\mu \nu}\right)N \nonumber \\
&& \pm\, ie\frac{g_{RN\gamma}^{(2)}}{\left(2M_N\right)^3}\bar{R}_{\mu \alpha} \Gamma^{(\mp)}\left(\partial^\alpha F^{\mu \nu}\right)\partial_\nu N \nonumber \\
&& + \, \hc,  \\[6pt]
{\cal L}_{RN\gamma}^{7/2\pm} &=&  ie\frac{g_{RN\gamma}^{(1)}}{\left(2M_N\right)^3}\bar{R}_{\mu \alpha \beta}\gamma_\nu \Gamma^{(\pm)}\left(\partial^{\alpha}\partial^{\beta} F^{\mu \nu}\right)N \nonumber \\
&&-\, e\frac{g_{RN\gamma}^{(2)}}{\left(2M_N\right)^4}\bar{R}_{\mu \alpha \beta} \Gamma^{(\pm)} \left(\partial^\alpha \partial^\beta F^{\mu \nu}\right) \partial_\nu N  \nonumber \\
&&  + \, \hc,
\end{eqnarray}
where $R$ designates the nucleon resonance, and the superscript of ${\cal L}_{RN\gamma}$ denotes the spin and parity of the resonance $R$. The coupling constants $g_{RN\gamma}^{(i)}$ $(i=1,2)$ are treated as fit parameters.

The effective Lagrangians for hadronic vertices including nucleon resonances read
\begin{eqnarray}
{\cal L}_{RN f_1}^{1/2\pm} &=& - \frac{g_{RN f_1}}{2M_N}\bar{R}\Gamma^{(\pm)} \left[ \left(\frac{\gamma_\mu\partial^2}{M_R\mp M_N} \pm i\partial_\mu \right) f_1^\mu\right] N \nonumber \\
&&  +\, \hc,   \\[6pt]
{\cal L}_{RN f_1}^{3/2\pm} &=&  i \frac{g_{RN f_1}}{2M_N}\bar{R}_\mu \gamma_\nu \Gamma^{(\mp)}f_1^{\mu \nu} N  + \hc,  \\[6pt]
{\cal L}_{RN f_1}^{5/2\pm} &=& - \frac{g_{RN f_1}}{\left(2M_N\right)^2}\bar{R}_{\mu \alpha}\gamma_\nu \Gamma^{(\pm)}\left(\partial^{\alpha} f_1^{\mu \nu}\right) N  + \hc,  \\[6pt]
 {\cal L}_{RN f_1}^{7/2\pm} &=& - i \frac{g_{RN f_1}}{\left(2M_N\right)^3}\bar{R}_{\mu \alpha \beta}\gamma_\nu \Gamma^{(\mp)} \left(\partial^{\alpha}\partial^{\beta} f_1^{\mu \nu}\right) N \nonumber \\
 && + \, \hc
\end{eqnarray}
The couplings $g_{RN f_1}$ are fit parameters. Actually, in a single-channel calculation as the one performed in the present work, only the products of the electromagnetic coupling constants and hadronic coupling constants are relevant to the reaction amplitudes, and in practice we fit these products directly instead of fitting the hadronic and electromagnetic couplings separately.

\subsection{Resonance propagators}

In the present work, contributions from $s$-channel pole diagrams of a minimal number of nucleon resonances with various spin and parity are included to describe the data. For spin-$1/2$ resonance, the propagator reads
\begin{equation}
S_{1/2}(p) = \frac{i}{\slashed{p} - M_R + i \Gamma/2},
\end{equation}
where $M_R$, $\Gamma$, and $p$ are mass, width, and four-momentum of resonance $R$, respectively.

Following Refs.~\cite{Behrends:1957,Fronsdal:1958,Zhu:1999}, the prescriptions of the propagators for resonances with spin-$3/2$, -$5/2$, and -$7/2$ read
\begin{eqnarray}
S_{3/2}(p) &=&  \frac{i}{\slashed{p} - M_R + i \Gamma/2} \left( \tilde{g}_{\mu \nu} + \frac{1}{3} \tilde{\gamma}_\mu \tilde{\gamma}_\nu \right),  \\[6pt]
S_{5/2}(p) &=&  \frac{i}{\slashed{p} - M_R + i \Gamma/2} \,\bigg[ \, \frac{1}{2} \big(\tilde{g}_{\mu \alpha} \tilde{g}_{\nu \beta} + \tilde{g}_{\mu \beta} \tilde{g}_{\nu \alpha} \big)  \nonumber \\
&& -\, \frac{1}{5}\tilde{g}_{\mu \nu}\tilde{g}_{\alpha \beta}  + \frac{1}{10} \big(\tilde{g}_{\mu \alpha}\tilde{\gamma}_{\nu} \tilde{\gamma}_{\beta} + \tilde{g}_{\mu \beta}\tilde{\gamma}_{\nu} \tilde{\gamma}_{\alpha}  \nonumber \\
&& +\, \tilde{g}_{\nu \alpha}\tilde{\gamma}_{\mu} \tilde{\gamma}_{\beta} +\tilde{g}_{\nu \beta}\tilde{\gamma}_{\mu} \tilde{\gamma}_{\alpha} \big) \bigg], \\[6pt]
S_{7/2}(p) &=&  \frac{i}{\slashed{p} - M_R + i \Gamma/2} \, \frac{1}{36}\sum_{P_{\mu} P_{\nu}} \bigg( \tilde{g}_{\mu_1 \nu_1}\tilde{g}_{\mu_2 \nu_2}\tilde{g}_{\mu_3 \nu_3} \nonumber \\
&& -\, \frac{3}{7}\tilde{g}_{\mu_1 \mu_2}\tilde{g}_{\nu_1 \nu_2}\tilde{g}_{\mu_3 \nu_3} + \frac{3}{7}\tilde{\gamma}_{\mu_1} \tilde{\gamma}_{\nu_1} \tilde{g}_{\mu_2 \nu_2}\tilde{g}_{\mu_3 \nu_3} \nonumber \\
&& -\, \frac{3}{35}\tilde{\gamma}_{\mu_1} \tilde{\gamma}_{\nu_1} \tilde{g}_{\mu_2 \mu_3}\tilde{g}_{\nu_2 \nu_3} \bigg),  \label{propagator-7hf}
\end{eqnarray}
where
\begin{eqnarray}
\tilde{g}_{\mu \nu} &=& -\, g_{\mu \nu} + \frac{p_{\mu} p_{\nu}}{M_R^2}, \\[6pt]
\tilde{\gamma}_{\mu} &=& \gamma^{\nu} \tilde{g}_{\nu \mu} = -\gamma_{\mu} + \frac{p_{\mu}\slashed{p}}{M_R^2}.
\end{eqnarray}

\subsection{Form factors}

Each hadronic vertex obtained from the Lagrangians given in Sec.~\ref{Sec:Lagrangians} is accompanied with a phenomenological form factor to parametrize the structure of the hadrons and to normalize the behavior of the production amplitude. Following Refs.~\cite{Wang:2017,Wang:2018}, for intermediate baryon exchange we take the form factor as
\begin{equation}
f_B(p^2) = \left[\frac{\Lambda_B^4}{\Lambda_B^4 + \left(p^2-M_B^2\right)^2}\right]^2,  \label{eq:ff_B}
\end{equation}
where $p$ denotes the four-momentum of the intermediate baryon and $M_B$ is the mass for exchanged baryon $B$. For intermediate meson exchange, we take the form factor as
\begin{equation}
f_M(q^2) = \frac{\Lambda_M^2-M_M^2}{\Lambda_M^2-q^2}, \label{eq:ff_M}
\end{equation}
where $q$ represents the four-momentum of the intermediate meson, and $M_M$ is the mass of exchanged meson $M$. The cutoffs $\Lambda_{B(M)}$ for each exchanged baryon (meson) in Eqs.~\eqref{eq:ff_B} and \eqref{eq:ff_M} are treated as fit parameters.

\subsection{Treatments of $t$-channel reaction amplitudes}  \label{Sec:Regge_treatments}

The $t$-channel reaction amplitudes are usually constructed in three different types, i.e., the Feynman type, the Regge type, and the interpolated Regge type \cite{Wang:2020}. In the present work, we try all these three possibilities to explore how the reaction mechanisms of $\gamma p \to f_1(1285) p$ and the extracted resonance contents and parameters depend on the choices of different types of $t$-channel interactions, and what can we really learn from the available cross-section data for $\gamma p \to f_1(1285) p$.

\subsubsection{Feynman model}

In Feynman model, the reaction amplitudes from $t$-channel $\rho$ and $\omega$ exchanges are constructed directly by evaluating the Feynman diagram of Fig.~\ref{FIG:feymans}(b). The electromagnetic and hadronic vertices can be obtained directly by use of the Lagrangians given in Sec.~\ref{Sec:Lagrangians}. The amplitudes for $\rho$ and $\omega$ exchanges read
\begin{align}
{\cal M}_V^{\nu\mu} = & - i \, g_{NNV} \, g_{Vf_1\gamma} \, \epsilon^{\alpha\beta\mu\nu} \, k_\alpha \, t \,\frac{-g^{\beta\eta} + q^\beta q^\eta / m_V^2}{t-m_V^2} \nonumber \\
& \times \left[\gamma^\eta - i\frac{\kappa_V}{2M_N} \sigma^{\eta\delta} q_\delta \right],
\end{align}
where $V$ represents $\rho$ or $\omega$, and $q$ denotes the four-momentum of the exchanged vector meson.

\subsubsection{Regge model} \label{subsec:Regge}

In high energy region, the differential cross sections are dominated at forward angles, where the effects of high-spin meson exchanges matter. An economic way to describe the $t$- and $s$-dependence of the cross sections in forward direction is Regge phenomenology \cite{Huang:2009,Sibirtsev:2009,Huang:2010}. The standard Reggeization of the $t$-channel Feynman amplitudes of $\rho$ and $\omega$ exchanges corresponds to the following replacements of the propagators:
\begin{align}
\frac{1}{t-m^2_V}  \quad\Longrightarrow\quad  {\cal P}^V_{\rm R} = & \; \left(\frac{s}{s_0}\right)^{\alpha_V(t)-1}  \frac{\pi \alpha'_V}{\sin[\pi\alpha_V(t)]}  \nonumber \\[3pt]
& \; \times \frac{1}{\Gamma[\alpha_V(t)]}.  \label{eq:Regge_prop_meson}
\end{align}
Here $s_0$ is a mass scale which is conventionally taken as $s_0=1$ GeV$^2$, and $\alpha'_V$ is the slope of the Regge trajectory $\alpha_V(t)$. For $V=\rho$ and $\omega$, the trajectories are parametrized as \cite{Guidal:1997}
\begin{align}
\alpha_{\rho}(t) &= 0.55 + 0.8~{\rm GeV}^{-2} t, \label{eq:trajectory_rho}  \\[6pt]
\alpha_{\omega}(t) &= 0.44 + 0.9~{\rm GeV}^{-2}t . \label{eq:trajectory_omega}
\end{align}
Note that in Eq.~\eqref{eq:Regge_prop_meson} the degenerate trajectories are employed for $\rho$ and $\omega$ exchanges, thus the corresponding signature factors reduce to $1$. Such a choice is preferred by data, which has been tested by our numerical calculation.

\subsubsection{Interpolated Regge model} \label{subsec:interpolated_Regge}

It is believed that Regge model works properly in the large-$s$ and small-$|t|$ region, and Feynman model works properly in the low-energy region. In literature, the so-called interpolated Regge model is widely used \cite{Toki:2008,Nam:2010,He:2012,Wang:2020}. In this model, an interpolating form factor is introduced to parametrize the smooth transition from Feynman amplitudes to Regge amplitudes. Instead of the replacement given in Eq.~\eqref{eq:Regge_prop_meson}, in the interpolated Regge model, one has the following replacement of the propagators of $t$-channel $\rho$ and $\omega$ exchanges in Feynman amplitudes:
\begin{align}
\frac{1}{t-m^2_V}  \quad\Longrightarrow\quad  {\cal P}^V_{\rm IR} = {\cal P}^V_{\rm R} F + \frac{1}{t-m^2_V} \left(1-F\right),   \label{Eq: intRe}
\end{align}
where $F = F_s F_t$ with
\begin{align}
F_s &= \frac{1}{1+e^{-\left(s-s_R\right)/s_0}},  \label{eq:IR_Fs}  \\[6pt]
F_t &= \frac{1}{1+e^{-\left(t+t_R\right)/t_0}}.   \label{eq:IR_Ft}
\end{align}
Here $s_R$, $s_0$, $t_R$, and $t_0$ are parameters to be fixed by fitting the data. The parameters $s_R$ and $t_R$ denote where do the amplitudes transit from Feynman type to Regge type, and $s_0$ and $t_0$ indicate how fast do this transit occur.

It is seen that by making the replacement of Eq.~\eqref{Eq: intRe}, the $t$-channel amplitude is a combination of Regge amplitude and Feynman amplitude, with a weight factor $F$ for the former and $(1-F)$ for the latter. In the low-energy $(s\ll s_R)$ region, the factor $F$ tends towards $0$, ensuring that one has almost pure Feynman amplitude. In the high-energy $(s\gg s_R)$ and small-$|t|$ $(|t|\ll t_R)$ region, the factor $(1-F)$ tends towards $0$, ensuring that one has almost pure Regge amplitude. In the intermediate energy region, the amplitude is constructed as a mixture of Feynman amplitude and Regge amplitude, which transits smoothly to Feynman amplitude in low-energy region and Regge amplitude in high-energy region.

\section{Results and discussion}   \label{Sec:results}

In the present work, we perform a comprehensive investigation of the $\gamma p \to f_1(1285) p$ reaction within an effective Lagrangian approach. The contributions from the $s$-channel $N$-pole diagram, $u$-channel $N$-exchange diagram, $t$-channel $\rho$- and $\omega$-exchange diagrams, and the interaction current are considered as background ingredients in constructing the reaction amplitudes. We build three reaction models, i.e., the Feynman model, the Regge model, and the interpolated Regge model, where the $t$-channel reaction amplitudes of $\rho$ and $\omega$ exchanges are built in Feynman type, Regge type, and interpolated Regge type, respectively (cf. Sec.~\ref{Sec:Regge_treatments}). In each model, we consider pole diagrams in the $s$ channel  of as few as possible nucleon resonances in order to achieve satisfactory descriptions of the available differential cross-section data.

In the most recent PDG review \cite{Workman:2022}, there are nine nucleon resonances with spin $J=1/2 \sim 7/2$ in the energy region considered in the present work, namely the $N(1990){7/2}^+$, $N(2000){5/2}^+$, $N(2040){3/2}^+$, $N(2060){5/2}^-$, $N(2100){1/2}^+$, $N(2120){3/2}^-$, $N(2190){7/2}^-$, $N(2300){1/2}^+$, and $N(2570){5/2}^-$ resonances. As it is not clear how many nucleon resonances and what nucleon resonances are really needed in the $\gamma p \to f_1(1285) p$ reaction, we follow the strategy employed in Refs.~\cite{Zhangy:2021,Wei:2022,Wang:2022,Wei:2020,Wangx:2020,Zhangyx:2021,Wein:2021,Weinc:2019} to choose nucleon resonances, i.e., we strive to describe the available data by introducing as few as possible nucleon resonances in our theoretical models. In practice, we make numerous trials of various number and different combination of these nine nucleon resonances in our calculations. Firstly, we try to reproduce the data without considering any nucleon resonance. If fail, we introduce one nucleon resonance to the model by testing the above-mentioned nine nucleon resonances one by one. If the data still cannot be well reproduced, various combinations of two nucleon resonances will be further considered in the model. In principle, more nucleon resonances should be further introduced in the calculations unless the data can be satisfactorily described. Nevertheless, for the $\gamma p \to f_1(1285)p$ reaction, so far, we only have differential cross-section data with relatively large error bars at $5$ energy points. Adding more than two nucleon resonances to the model will bring in too many parameters (masses, widths, cutoffs, and coupling constants, et al.) to the model which are difficult to be well determined by the available (very limited) data. Therefore, in each model we stop introducing more than two nucleon resonances if the data cannot be satisfactorily described.

In the following, we show and discuss the results from each model separately.

\subsection{Results in Feynman model} \label{subsec:results_f}

\begin{figure}[tbp]
\includegraphics[width=\columnwidth]{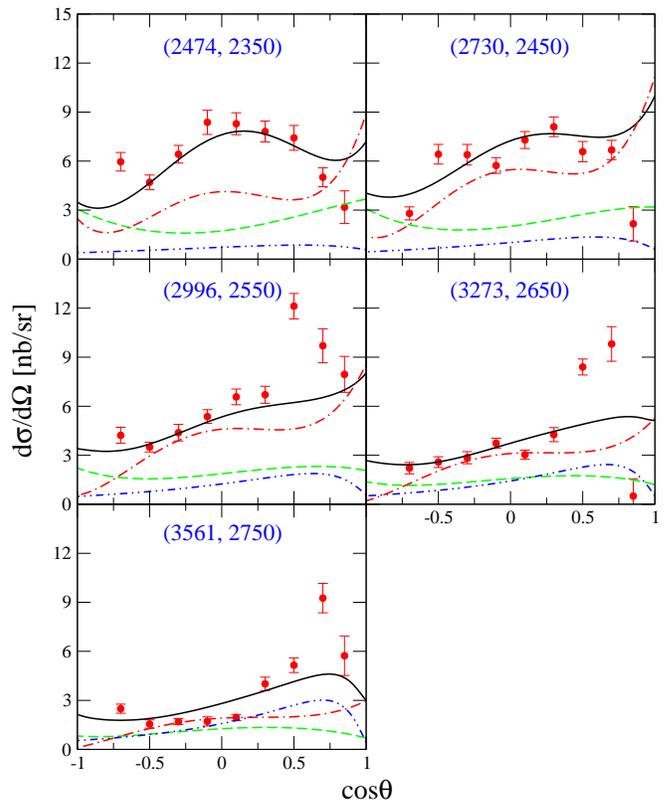}
\caption{Differential cross sections of $\gamma p \to f_1(1285)p$ as a function of cosine of scattering angle $\theta$ in center-of-mass frame obtained in the Feynman model with the contributions from the $N(2000){5/2}^+$ and $N(2060){5/2}^-$ resonances being considered. The black solid lines represent the results from the full calculation. The blue dash-double-dotted, green dashed, and red dash-dotted lines represent the individual contributions from $t$-channel $\rho$- and $\omega$-exchange diagrams, $s$-channel $N(2060){5/2}^-$-pole diagram, and $s$-channel $N(2000){5/2}^+$-pole diagram, respectively. The scattered symbols denote the CLAS data in Ref.~\cite{Dickson:2016gwc}. The numbers in parentheses denote the centroid value of the photon laboratory incident energy (left number) and the corresponding total center-of-mass energy of the system (right number), in MeV.}
\label{fig:dif_f_2res}
\end{figure}

\begin{table}[tb]
\caption{\label{Table:para-F} Fitted values of parameters in Feynman model with the $N(2000){5/2}^+$ and $N(2060){5/2}^-$ resonances being considered. The asterisks below resonance names denote the overall status of these resonances evaluated by the PDG review \cite{Workman:2022}. The numbers in brackets below resonance mass $M_R$ and width $\Gamma_R$ represent the range of the corresponding quantities given by the PDG review \cite{Workman:2022}.}
\begin{tabular*}{\columnwidth}{@{\extracolsep\fill}lcc}
\hline\hline
$\kappa_{f_1}$           & $-5.96\pm 5.0$            \\
$\Lambda_{N}$ [MeV]        & $578\pm8$             \\
$\Lambda_{\rho}$ [MeV]      & $769\pm1 $          \\
$\Lambda_{\omega}$ [MeV]    & $834\pm3$            \\
\hline
          &  $N(2000){5/2}^+$        &  $N(2060){5/2}^-$         \\
                               & $\ast$$\ast$    & $\ast$$\ast$$\ast$      \\
$M_R$ [MeV]                  &  $2250\pm3$      &  $2030\pm7$           \\
                                & [$\sim2000$]       & [$2030\sim2200$]              \\
$\Gamma_R$ [MeV]               &  $126\pm30$    &  $450\pm52$           \\
                              &  [$\sim300$]         &  [$\sim400$]                   \\
$\Lambda_R$ [MeV]              &  $1396\pm 28$       &  $1399\pm 59$          \\
$g_{RN\gamma}^{(1)} g_{RNf_1}$     &  $0.013\pm0.002$  &  $-18.5\pm0.8$        \\
$g_{RN\gamma}^{(2)} g_{RNf_1}$     &  $16.8\pm0.7$  &  $-61.0\pm2.0$     \\
\hline\hline
\end{tabular*}
\end{table}

In the Feynman model, it is expected that the near threshold structures exhibited in the angular distributions of $\gamma p \to f_1(1285)p$ are dominated by contributions from nucleon resonances, and the cross-section peaks in the high-energy region at forward angles are governed by contributions from $t$-channel $\rho$ and $\omega$ exchanges. However, after numerous tests, we found that in the cases when no nucleon resonance or any one or any two of the above-mentioned nine nucleon resonances are introduced in the calculations, the available differential cross-section data for $\gamma p \to f_1(1285)p$ from CLAS Collaboration \cite{Dickson:2016gwc} cannot be well reproduced in the Feynman model.

In Fig.~\ref{fig:dif_f_2res}, we show the results of differential cross sections of $\gamma p \to f_1(1285)p$  as a function of cosine of scattering angle $\theta$ in center-of-mass frame obtained in the Feynman model with the $N(2000){5/2}^+$ and $N(2060){5/2}^-$ contributions being considered. The fit with these two nucleon resonances being considered is the best one as the corresponding $\chi^2/N$ ($\chi^2$ per data point) is the smallest among all the fits with no nucleon resonance, one nucleon resonance, or two nucleon resonances being considered. In Fig.~\ref{fig:dif_f_2res}, the black solid lines represent the results from the full calculation. The blue dash-double-dotted, green dashed, and red dash-dotted lines represent the individual contributions from $t$-channel $\rho$- and $\omega$-exchange diagrams, $s$-channel $N(2060){5/2}^-$-pole diagram, and $s$-channel $N(2000){5/2}^+$-pole diagram, respectively. The scattered symbols denote the CLAS data from Ref.~\cite{Dickson:2016gwc}. The numbers in parentheses denote the centroid value of the photon laboratory incident energy (left number) and the corresponding total center-of-mass energy of the system (right number), in MeV. One sees from Fig.~\ref{fig:dif_f_2res} that the differential cross-section data in low-energy region can be well described, while the data in high-energy region near the forward angles are too much underestimated. Therefore, this set of results is not considered as acceptable. The model parameters corresponding to this set of theoretical results are listed in Table~\ref{Table:para-F}. One sees that the fitted mass and width of $N(2000){5/2}^+$ are a little bit far away from the corresponding values advocated in the PDG review \cite{Workman:2022}. If we manually restrict these parameters in a range as advocated in the PDG review \cite{Workman:2022} in the fitting procedure, the agreement of the fitted differential cross sections with the corresponding data will be even worse than those shown in Fig.~\ref{fig:dif_f_2res}.

The reason that the differential cross-section data of $\gamma p \to f_1(1285)p$ in high-energy region near forward angles cannot be well reproduced in the Feynman model is possibly that the Feynman type amplitudes of $t$-channel $\rho$ and $\omega$ exchanges do not result in proper angular dependence of the high-energy differential cross sections as demonstrated by the data. As so far we only have differential cross-section data at $5$ energy points, introducing more nucleon resonances to this model to improve the quality of the theoretical description of the data does not make too much sense. We thus conclude that the available differential cross-section data of $\gamma p \to f_1(1285)p$ cannot be well reproduced in Feynman model.

\subsection{Results in Regge model}

\begin{table}[tb]
\caption{\label{Table:chi2} $\chi^2/N$ for differential cross sections fitted by including one nucleon resonance in Regge model. The asterisks below resonance names denote the overall status of these resonances rated by the PDG review \cite{Workman:2022}.}
\begin{tabular*}{\columnwidth}{@{\extracolsep\fill}lccc}
\hline\hline
$N^\ast$   &  $N(1990){7/2}^+$  & $N(2000){5/2}^+$  &  $N(2040){3/2}^+$   \\
                 &   $\ast$$\ast$    &  $\ast$$\ast$    &  $\ast$    \\
$\chi^2/N$ & $6.5$ & $4.0$ & $4.1$   \\
\hline
$N^\ast$   &  $N(2060){5/2}^-$ & $N(2100){1/2}^+$  &  $N(2120){3/2}^-$   \\
                &  $\ast$$\ast$$\ast$ & $\ast$$\ast$$\ast$        &  $\ast$$\ast$$\ast$    \\
$\chi^2/N$  & $5.1$ & $6.5$ & $4.9$   \\
\hline
$N^\ast$     &  $N(2190){7/2}^-$  &  $N(2300){1/2}^+$ &  $N(2570){5/2}^-$  \\
                   &  $\ast$$\ast$$\ast$$\ast$     &  $\ast$$\ast$  & $\ast$$\ast$   \\
$\chi^2/N$   & $6.2$  & $4.7$  &  $4.6$ \\
\hline\hline
\end{tabular*}
\end{table}

\begin{figure}[tbp]
\includegraphics[width=\columnwidth]{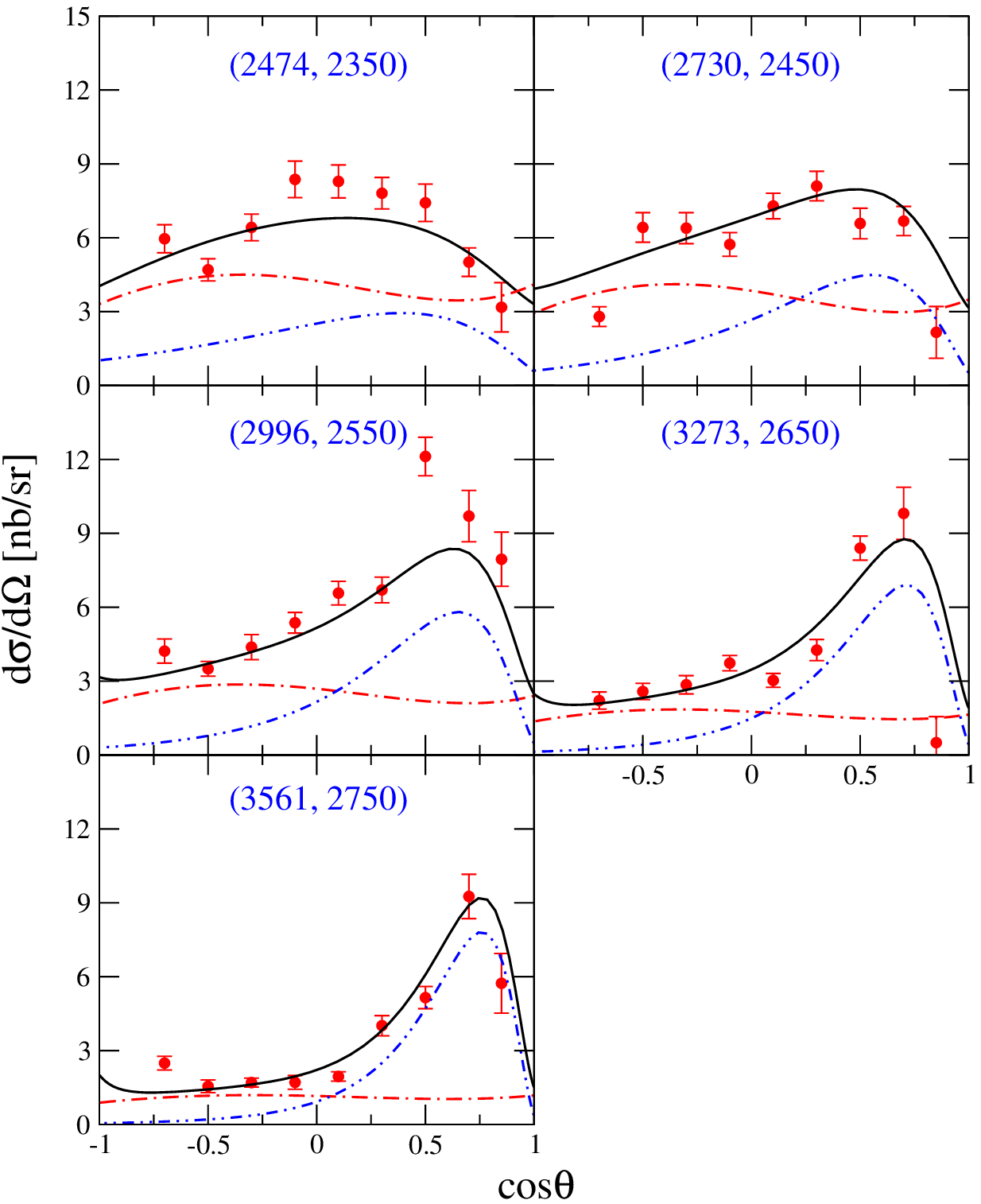}
\caption{Differential cross sections of $\gamma p \to f_1(1285)p$ as a function of cosine of scattering angle $\theta$ in center-of-mass frame obtained in the Regge model with the contribution from $s$-channel $N(2000)5/2^+$-pole diagram being considered. Notations are the same as in Fig.~\ref{fig:dif_f_2res}.}
\label{fig:dif_5}
\end{figure}

\begin{table}[tb]
\caption{\label{Table:para-R-IR} Fitted values of parameters in Regge model (second column) and interpolated Regge model (third column) with the $N(2000){5/2}^+$ resonance being considered. The asterisks below resonance names denote the overall status of these resonances evaluated by the PDG review \cite{Workman:2022}. The numbers in brackets below resonance mass $M_R$ and width $\Gamma_R$ represent the range of the corresponding quantities given by the PDG review \cite{Workman:2022}.}
\begin{tabular*}{\columnwidth}{@{\extracolsep\fill}lcc}
\hline\hline
                          &   Regge   &   interpolated Regge  \\
\hline
$\kappa_{f_1}$        &$14.4\pm 7.0$    & $14.6\pm 5.0$            \\
$\Lambda_{N}$ [MeV]       &$500\pm 103$    & $500\pm 124$             \\
$\Lambda_{\rho}$ [MeV]    &$1083\pm 17$     & $1079\pm 5$          \\
$\Lambda_{\omega}$ [MeV]  &$745\pm 83$     & $749\pm 51$            \\
\hline
                          &  $N(2000){5/2}^+$  &  $N(2000){5/2}^+$        \\
                          &  $\ast$$\ast$     & $\ast$$\ast$             \\
$M_R$ [MeV]               &  $2042 \pm 33$     &  $2042 \pm 11$           \\
                          &  [$\sim 2000$]           & [$\sim 2000$]              \\
$\Gamma_R$ [MeV]          &  $450 \pm 57$      &  $450 \pm 72$            \\
                          &  [$\sim 300$]            &  [$\sim 300$]                 \\
$\Lambda_R$ [MeV]         & $1167 \pm 17$      &  $1172 \pm 3$           \\
$g_{RN\gamma}^{(1)} g_{RNf_1}$  & $-64.7 \pm 0.5$  &  $-62.8 \pm 0.5$       \\
$g_{RN\gamma}^{(2)} g_{RNf_1}$  &   $27.9\pm 1.5$  &  $28.6 \pm 1.5$       \\
\hline\hline
\end{tabular*}
\end{table}

\begin{figure}[tbp]
\includegraphics[width=\columnwidth]{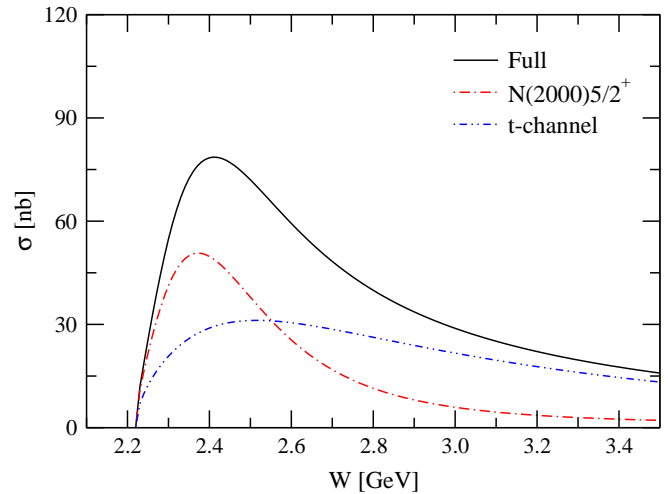}
\caption{Predicted total cross sections with dominant individual contributions for $\gamma p \to f_1(1285)p$ in Regge model with the contribution from $s$-channel $N(2000){5/2}^+$-pole diagram being considered. Notations are the same as in Fig.~\ref{fig:dif_5}.}
\label{fig:sig_5}
\end{figure}

\begin{figure}[tb]
\includegraphics[width=\columnwidth]{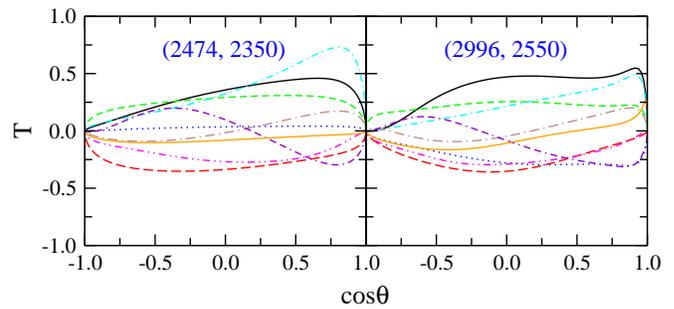}
\caption{Predicted target nucleon asymmetries as functions of $\cos\theta$ for $\gamma p \to f_1(1285)p$ at two selected energies. The numbers in parentheses denote the photon laboratory incident energy (left number) and the total center-of-mass energy of the system (right number), in MeV. The black solid, blue dotted, green dashed, red long-dashed, cyan dot-dashed, brown dot-long-dashed, magenta dash-double-dotted, violet dot-double-dashed, and orange solid curves represent the results in Regge model with consideration of the contribution from $s$-channel $N(2100)1/2^+$, $N(2300)1/2^+$, $N(2040)3/2^+$, $N(2120)3/2^-$, $N(2000)5/2^+$, $N(2060)5/2^-$, $N(1990)7/2^+$, $N(2190)7/2^-$, and $N(2570){5/2}^-$, respectively.}
\label{fig:target_asy}
\end{figure}

As indicated in Sec.~\ref{subsec:Regge}, the Regge type amplitudes of $t$-channel $\rho$ and $\omega$ exchanges have rather different angular dependence than the Feynman type amplitudes. Thus, although the Feynman model fails, the Regge model may still have the opportunity to describe the available differential cross-section data for $\gamma p \to f_1(1285)p$.

We made numerous tests and found that if no contribution from $s$-channel $N^\ast$-pole diagram is considered in the Regge model, the resulted $\chi^2/N$ will be greater than $10$, indicating significantly poor fitting quality of the obtained results. This means that in Regge model when no nucleon resonance is considered, the shapes of angular distributions in low-energy region and high-energy region cannot be simultaneously described by non-resonant contributions, which are mainly the $t$-channel $\rho$ and $\omega$ exchanges, only. We then tried to add one of those nine nucleon resonances as mentioned in the text above Sec.~\ref{subsec:results_f} to this model. We tested one by one, and it was found that by adding any one of those nine nucleon resonances in the Regge model, the available differential cross-section data of $\gamma p \to f_1(1285)p$ can always be satisfactorily described with similar fitting qualities, $4.0 < \chi^2/N < 6.5$, as listed in Table~\ref{Table:chi2}.

In Fig.~\ref{fig:dif_5}, we show the theoretical results, compared with the corresponding data, for differential cross sections of $\gamma p \to f_1(1285)p$ as a function of cosine of scattering angle $\theta$ in center-of-mass frame obtained in the Regge model with the contribution from $s$-channel $N(2000)5/2^+$-pole diagram being considered. The notations of this figure are the same as in Fig.~\ref{fig:dif_f_2res}. The fitted values of model parameters correspond to this set of results are listed in the second column of Table~\ref{Table:para-R-IR}. It is seen that the fitted resonance mass and width are consistent with the values advocated in the most recent PDG review \cite{Workman:2022}. Note that the resonance electromagnetic and hadronic couplings are not shown separately, since in a single-channel calculation as the one performed in the present work, the reaction amplitudes depend only on the products of them. In PDG review \cite{Workman:2022}, no electromagnetic decay amplitudes or $f_1(1285)p$ hadronic branching ratio for the two-star resonance $N(2000){5/2}^+$ are advocated. If we use the helicity amplitudes $A_{1/2}=0.031$ GeV$^{-1/2}$ and $A_{3/2}=-0.043$ GeV$^{-1/2}$ from a BnGa partial wave analysis \cite{Sokhoyan:2015} to fix the electromagnetic couplings, we get $g_{RN\gamma}^{(1)}=-3.9$, $g_{RN\gamma}^{(2)}=3.6$. A refit of the hadronic coupling results in $g_{RNf_1}=19.8$, while the cross-section results change very little.

One sees from Fig.~\ref{fig:dif_5} that the angular distribution data for $\gamma p \to f_1(1285)p$ in the whole energy region considered are quite well described in the Regge model with the $N(2000)5/2^+$ resonance being considered. In the high-energy region, the peaks at forward angles are dominated by the $t$-channel $\rho$ and $\omega$ Regge exchanges. In the low-energy region, the resonance contribution also provides significant contributions to the observed cross sections. Note that compared with the contributions from the $t$-channel $\rho$- and $\omega$-exchange diagrams in the Feynman model (cf. Fig.~\ref{fig:dif_f_2res}), now the contributions from the $t$-channel $\rho$- and $\omega$-trajectory exchanges in the Regge model provide much steeper uplifts of the angular distributions at forward angles, which is the main reason why the Regge model succeeds in reproducing the data while the Feynman model fails. Considering that a meson trajectory exchange is an economic treatment of the exchanges of all mesons lying on the trajectory, such a fact may indicate that for the $\gamma p \to f_1(1285)p$ reaction, contributions from the $t$-channel exchanges of high-spin mesons are important. The $\chi^2/N$ is $4.0$ for this set of solution. For the other eight sets of solutions, each with one of the other eight nucleon resonances being considered, the resulted $\chi^2/N$ are between $4.0$ and $6.5$. They all reproduce the available differential cross-section data for $\gamma p \to f_1(1285)p$ satisfactorily. The reaction mechanisms in those sets of solutions are shown to be quite similar --- the $t$-channel $\rho$ and $\omega$ exchanges dominate the high-energy peaks at forward angles, and the $s$-channel resonance contribution provides considerable contributions in low-energy region. Nevertheless, the available differential cross-section data are not sufficient to distinguish those different sets of solutions in the Regge model. 

The predicted total cross sections with dominant individual contributions for $\gamma p \to f_1(1285)p$ in Regge model with the $N(2000){5/2}^+$ resonance being considered are shown in Fig.~\ref{fig:sig_5}.   One sees that in the high-energy region, it is the $t$-channel interaction that dominates the cross sections, while in the low-energy region, the $N(2000){5/2}^+$ resonance also contributes significantly. This observation is consistent with the behavior of the differential cross sections as shown in Fig.~\ref{fig:dif_5}. The results for models with other nucleon resonances are very close to those with the $N(2000){5/2}^+$ resonance.

In Fig.~\ref{fig:target_asy} we show the predictions of target nucleon asymmetries at two selected energies in Regge model with each of the $N(2100){1/2}^+$, $N(2300){1/2}^+$, $N(2040){3/2}^+$, $N(2120){3/2}^-$, $N(2000){5/2}^+$, $N(2060){5/2}^-$, $N(1990){7/2}^+$, $N(2190){7/2}^-$, and $N(2570){5/2}^-$ resonances. One sees that the signs and magnitudes of the target nucleon asymmetries predicted in Regge model with various nucleon resonances are different. Future data on this quantity are expected to be used to distinguish the Regge models with various nucleon resonances.

\subsection{Results in interpolated Regge model}

\begin{figure}[tbp]
\includegraphics[width=\columnwidth]{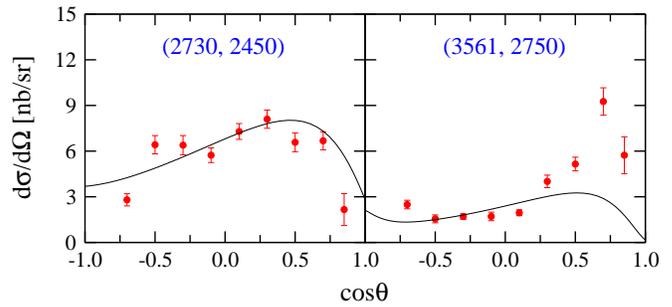}
\caption{Differential cross sections of $\gamma p \to f_1(1285)p$ as a function of cosine of scattering angle $\theta$ in center-of-mass frame obtained in the interpolated Regge model with no resonance being considered. Notations are the same as in Fig.~\ref{fig:dif_f_2res}.}
\label{fig:dif_interp_Regge_no_res}
\end{figure}

\begin{table}[tb]
\caption{\label{Table:para2} Fitted values of auxiliary parameters introduced in interpolated Regge model. See Sec.~\ref{subsec:interpolated_Regge} for the definitions of these parameters. All numbers are in GeV$^2$.}
\renewcommand{\arraystretch}{1.2}
\begin{tabular*}{\columnwidth}{@{\extracolsep\fill}cccc}
\hline\hline
$s_0$  & $s_R$  & $t_0$  & $t_R$  \\
$0.26\pm 0.35$ & $0.80 \pm 2.30$ & $0.05\pm 0.70$ & $10.20\pm 6.25$ \\
\hline\hline
\end{tabular*}
\end{table}

As mentioned in the introduction section, two works in literature \cite{Wangxiaoyun:2017,Wangyanyan:2017}  have already been devoted to the study of $\gamma p \to f_1(1285)p$ photoproduction reaction in the interpolated Regge model. Although these two theoretical works used the same reaction model, their results and conclusions are quite different. In Ref.~\cite{Wangxiaoyun:2017}, it was claimed that the available differential cross-section data for $\gamma p \to f_1(1285)p$ can be reproduced without considering any nucleon resonances. However, in Ref.~\cite{Wangyanyan:2017}, it was argued that the $N(2300){1/2}^+$ resonance plays an important role in describing the available differential cross-section data for $\gamma p \to f_1(1285)p$. In the present work, we also perform a comprehensive investigation of $\gamma p \to f_1(1285)p$ in interpolated Regge model, with special attention paying to whether the available data for $\gamma p \to f_1(1285)p$ can be reproduced in interpolated Regge model without or with the contributions from any nucleon resonances.

As the interpolated Regge model contains four more parameters compared with the Regge model [cf. Eqs.~\eqref{eq:IR_Fs} and \eqref{eq:IR_Ft} in Sec.~\ref{subsec:interpolated_Regge}], naively, one may expect that the interpolated Regge model can describe the data better than the Regge model. However, this is not the truth.

Firstly, we varied the parameters in non-resonant contributions by use of MINUIT to fit the data in the interpolated Regge model without considering any nucleon resonances. After numerous trials, it was found that the fitting quality of the results is poor, as the $\chi^2/N$ is $7.9$. The fitted results are illustrated in Fig.~\ref{fig:dif_interp_Regge_no_res}, from which one sees clearly that the available differential cross-section data for $\gamma p \to f_1(1285)p$ cannot be reasonably described in the interpolated Regge model when no $s$-channel $N^\ast$-pole diagrams are considered. Such a conclusion contradicts with Ref.~\cite{Wangxiaoyun:2017} but is in agreement with Ref.~\cite{Wangyanyan:2017}. Note that in Ref.~\cite{Wangxiaoyun:2017}, the contributions from the $NN\gamma$ vector coupling [$\propto {\bar N}\gamma_\mu A^\mu N$, cf. Eq.~\eqref{eq:L_NNg}] in $s$-channel $N$-pole diagram and $u$-channel $N$-exchange diagram, and the contact term [cf. Eqs.~\eqref{eq:amplitude} and \eqref{eq:Mint}] were totally omitted, which sounds need further justification.

We then tried to add one of those nine nucleon resonances as mentioned in the text above Sec.~\ref{subsec:results_f} to the interpolated Regge model to reproduce the data. Surprisingly, for each set of results, the obtained $\chi^2/N$ is very close to that of the solution in the Regge model with the same resonance content. Actually, the corresponding values of $\chi^2/N$ in these two models only differ in the second digits (see Table~\ref{Table:chi2} for $\chi^2/N$ in Regge model). Moreover, the fitted values of the model parameters in the interpolated Regge model are also very close to those in the Regge model when the same nucleon resonance is considered. As an example, the fitted values of the model parameters with the $N(2000){5/2}^+$ resonance considered in the interpolated Regge model are listed in the third column of Table~\ref{Table:para-R-IR}. One sees that they are very close to those listed in the second column of this table for the Regge model.

This unusual phenomenon is not difficult to be understood from the fitted values of those four auxiliary parameters introduced in the interpolated Regge model, as listed in Table~\ref{Table:para2}. The parameter $s_R$ indicates that around which energy point does the Feynman amplitude transit to the Regge amplitude, and the parameter $s_0$ demonstrates how fast does this transition occur. From Table~\ref{Table:para2} one sees that for the $\gamma p \to f_1(1285)p$ reaction, the $t$-channel amplitudes transit from Feynman type to Regge type around $s_R=0.8$ GeV$^2$, and the transition occurs very fast as $s_0$ is a very small number, $0.26$ GeV$^2$. This means that above the $f_1(1285)p$ threshold, i.e., $s\sim 4.95$ GeV$^2$, the $t$-channel amplitudes are almost purely in Regge type. In other words, the so-called interpolated Regge model is in substance the Regge model for the $\gamma p \to f_1(1285)p$ reaction. We observe that in Ref.~\cite{Wangyanyan:2017}, the values of these two parameters also manifest that the interpolated Regge model claimed by the authors is essentially the Regge model.

If we manually force the parameter $s_R$ greater than the $f_1(1285)p$ threshold value $\sim 4.95$ GeV$^2$ and $s_0$ not that small in the fitting procedure, so that the reaction amplitudes indeed have the spirits of interpolated Regge model, i.e., they transit smoothly from Feynman type in low-energy region to Regge type in high-energy region, the resulted $\chi^2/N$ will be greater than $10.0$, and consequently the fitting quality is too inferior to be accepted. We thus conclude that the available angular distribution data for $\gamma p \to f_1(1285)p$ cannot be reproduced in interpolated Regge model.

\section{Summary and conclusion}  \label{sec:summary}

The CLAS differential cross-section data for $\gamma p \to f_1(1285) p$ reaction \cite{Dickson:2016gwc} have so far been analyzed in Refs.~\cite{Wangxiaoyun:2017,Wangyanyan:2017}. Although both of these two theoretical works used the interpolated Regge model, their conclusions are quite different. In Ref.~\cite{Wangxiaoyun:2017}, it was claimed that the available data for $\gamma p \to f_1(1285) p$ can be reproduced in the interpolated Regge model without considering any nucleon resonances. However, in Ref.~\cite{Wangyanyan:2017}, it was argued that the $N(2300)1/2^+$ resonance is needed in the interpolated Regge model to describe the data.

In the present work, we performed a comprehensive investigation of the $\gamma p \to f_1(1285) p$ reaction in an effective Lagrangian approach. We considered the contributions from $s$-channel $N$-pole diagram, $u$-channel $N$-exchange diagram, $t$-channel $\rho$- and $\omega$-exchange diagrams, and generalized interaction current in constructing the non-resonant reaction amplitudes. We built three reaction models, i.e., the Feynman model, the Regge model, and the interpolated Regge model, where the $t$-channel amplitudes are constructed in Feynman type, Regge type, and interpolated Regge type, respectively. In each model, we introduced as few as possible nucleon resonances to reproduce the data. All nine nucleon resonances listed in the most recent PDG review \cite{Workman:2022} with spin $J=1/2 \sim 7/2$ in the energy region considered in the present work, namely the $N(1990){7/2}^+$, $N(2000){5/2}^+$, $N(2040){3/2}^+$, $N(2060){5/2}^-$, $N(2100){1/2}^+$, $N(2120){3/2}^-$, $N(2190){7/2}^-$, $N(2300){1/2}^+$, and $N(2570){5/2}^-$ resonances, are allowed in practice. The purpose of the present work was to find concise and clear answers of the following questions: i) Whether the available differential cross-section data for $\gamma p \to f_1(1285) p$ can be reproduced in Feynman model, Regge model, and interpolated Regge model, respectively? ii) If yes, how many nucleon resonances are at least needed and what are the resonance contents and associated parameters in these three models?

Our results showed that the available angular distribution data for $\gamma p \to f_1(1285) p$ from the CLAS Collaboration \cite{Dickson:2016gwc} cannot be well reproduced in either Feynman model or interpolated Regge model. In the Regge model, when no nucleon resonance is considered, the data cannot be satisfactorily described either. However, when any one of the $N(1990){7/2}^+$, $N(2000){5/2}^+$, $N(2040){3/2}^+$, $N(2060){5/2}^-$, $N(2100){1/2}^+$, $N(2120){3/2}^-$, $N(2190){7/2}^-$, $N(2300){1/2}^+$, and $N(2570){5/2}^-$ resonances is taken into account in the Regge model, the experimental data can be well described. The resulted resonance masses and widths are consistent with those advocated in PDG review \cite{Workman:2022}. In each set of results in the Regge model, it was found that in the high-energy region, the peaks of differential cross sections at forward angles are dominated by $t$-channel $\rho$ and $\omega$ exchanges, and in the low-energy region, the resonance contribution provides significant contributions also.

The fact that the available differential cross-section data for $\gamma p \to f_1(1285) p$ can only be reproduced in Regge model but not Feynman model or interpolated Regge model reveals that the angular and energy dependence of the $f_1(1285)$ photoproduction amplitudes is rather unusual. This indicates that the $f_1(1285)$ meson might have an unconventional $q{\bar q}$ content which deserves further investigations.

\begin{acknowledgments}
This work is partially supported by the National Natural Science Foundation of China under Grants No.~12175240, No.~12147153, and No.~11635009, the Fundamental Research Funds for the Central Universities, and the China Postdoctoral Science Foundation under Grants No.~2021M693141 and No.~2021M693142.
\end{acknowledgments}


\begin{thebibliography}{99}
%
\bibitem{Isgur:1978}
N. Isgur and G. Karl, Phys. Rev. D {\bf 18}, 4187 (1978).
%
\bibitem{Capstick:1986}
S. Capstick and N. Isgur, Phys. Rev. D {\bf 34}, 2809 (1986).
%
\bibitem{Loring:2001}
U. L\"{o}ring, B. C. Metsch, and H. R. Petry, Eur. Phys. J. A {\bf 10}, 395 (2001).
%
\bibitem{Workman:2022}
R. L. Workman {\it et al.} (Particle Data Group), Prog. Theor. Exp. Phys. {\bf 2022}, 083C01 (2022).
%
\bibitem{Kashevarov:2017}
V. L. Kashevarov {\it et al.} (CLAS Collaboration), Phys. Rev. Lett. {\bf 118}, 212001 (2017).
%
\bibitem{Anisovich:2018}
A. V. Anisovich, V. Burkert, P. M. Collins, M. Dugger, E. Klempt, V. A. Nikonov, B. G. Ritchie, A. V. Sarantsev, and U. Thoma, Phys. Lett. B {\bf 785}, 626 (2018).
%
\bibitem{Zhangy:2021}
Y. Zhang, A. C. Wang, N. C. Wei, and F. Huang, Phys. Rev. D {\bf 103}, 094036 (2021).
%
\bibitem{Bradford:2006}
R. Bradford {\it et al.} (CLAS Collaboration), Phys. Rev. C {\bf 73}, 035202 (2006).
%
\bibitem{Mart:2019}
T. Mart, Phys. Rev. D {\bf 100}, 056008 (2019).
%
\bibitem{CLAS-beam}
N. Zachariou \textit{et al.} (CLAS Collaboration), Phys. Lett. B \textbf{827}, 136985 (2022).
%
\bibitem{Zachariou:2020kkb}
N. Zachariou \textit{et al.} (CLAS Collaboration), Phys. Lett. B \textbf{808}, 135662 (2020).
\bibitem{Wei:2022}
N. C. Wei, A. C. Wang, F. Huang, and K. Nakayama, Phys. Rev. D {\bf 105}, 094017 (2022).
%
\bibitem{Wang:2022}
A. C. Wang, N. C. Wei, and F. Huang, Phys. Rev. D {\bf 105}, 034017 (2022).
%
\bibitem{Wei:2020}
N. C. Wei, A. C. Wang, F. Huang, and D. M. Li, Phys. Rev. C {\bf 101}, 014003 (2020).
%
\bibitem{Anisovich:2017rpe}
A. V. Anisovich {\it et al.} (CLAS Collaboration), Phys. Lett. B {\bf 771}, 142 (2017).
%
\bibitem{Moriya:2013}
K. Moriya {\it et al.} (CLAS Collaboration), Phys. Rev. C {\bf 88}, 045201 (2013).
%
\bibitem{Wangx:2020}
A. C. Wang, W. L. Wang, and F. Huang, Phys. Rev. D {\bf 101}, 074025 (2020).
%
\bibitem{Zhangyx:2021}
Y. Zhang and F. Huang, Phys. Rev. C {\bf 103}, 025207 (2021).
%
\bibitem{Wein:2021}
N. C. Wei, Y. Zhang, F. Huang, and D. M. Li, Phys. Rev. D {\bf 103}, 034007 (2021).
%
\bibitem{Weinc:2019}
N. C. Wei, F. Huang, K. Nakayama, and D. M. Li, Phys. Rev. D {\bf 100}, 114026 (2019).
%
\bibitem{Roy:2018}
P. Roy {\it et al.} (CLAS Collaboration), Phys. Rev. Lett. {\bf 122}, 162301 (2019).
%
\bibitem{Dickson:2016gwc}
R. Dickson {\it et al.} (CLAS Collaboration), Phys. Rev. C {\bf 93}, 065202 (2016).
%
\bibitem{Kochelev:2009}
N. I. Kochelev, M. Battaglieri, and R. DeVita, Phys. Rev. C {\bf 80}, 025201 (2009).
%
\bibitem{Domokos:2009}
S. K. Domokos, H. R. Grigoryan, and J. A. Harvey, Phys. Rev. D {\bf 80}, 115018 (2009).
%
\bibitem{HuangY:2014}
Y. Huang, J. J. Xie, X. R. Chen, J. He, and H. F. Zhang, Int. J. Mod. Phys. E {\bf 23}, 1460002 (2014).
%
\bibitem{Wangxiaoyun:2017}
X. Y. Wang and J. He, Phys. Rev. D {\bf 95}, 094005 (2017).
%
\bibitem{Wangyanyan:2017}
Y. Y. Wang, L. J. Liu, E. Wang, and D. M. Li, Phys. Rev. D {\bf 95}, 096015 (2017).
%
\bibitem{Haberzettl:1997}
H. Haberzettl, Phys. Rev. C {\bf 56}, 2041 (1997).
%
\bibitem{Haberzettl:2006}
H. Haberzettl, K. Nakayama, and S. Krewald, Phys. Rev. C {\bf 74}, 045202 (2006).
%
\bibitem{Huang:2012}
F. Huang, M. D\"{o}ring, H. Haberzettl, J. Haidenbauer, C. Hanhart, S. Krewald, U.-G. Mei{\ss}ner, and K. Nakayama, Phys. Rev. C {\bf 85}, 054003 (2012).
%
\bibitem{Huang:2013}
F. Huang, H. Haberzettl, and K. Nakayama, Phys. Rev. C {\bf 87}, 054004 (2013).
%
\bibitem{Drell:1972}
S. D. Drell and T. D. Lee, Phys. Rev. D {\bf 5}, 1738 (1972).
%
\bibitem{Birkel:1996}
M. Birkel and H. Fritzsch, Phys. Rev. D {\bf 53}, 6195 (1996).
%
\bibitem{Ronchen:2013}
D. R\"{o}nchen, M. D\"{o}ring, F. Huang, H. Haberzettl, J. Haidenbauer, C. Hanhart, S. Krewald, U.-G. Mei{\ss}ner, and K. Nakayama, Eur. Phys. J. A {\bf 49}, 44 (2013).
%
\bibitem{Behrends:1957}
R. E. Behrends and C. Fronsdal, Phys. Rev. {\bf 106}, 345 (1957).
%
\bibitem{Fronsdal:1958}
C. Fronsdal, Supp. Nuovo Cimento {\bf 9}, 416 (1958).
%
\bibitem{Zhu:1999}
J. J. Zhu and M. L. Yan, arXiv:hep-ph/9903349.
%
\bibitem{Wang:2017}
A. C. Wang, W. L. Wang, F. Huang, H. Haberzettl, and K. Nakayama, Phys. Rev. C {\bf 96}, 035206 (2017).
%
\bibitem{Wang:2018}
A. C. Wang, W. L. Wang, and F. Huang, Phys. Rev. C {\bf 98}, 045209 (2018).
%
\bibitem{Wang:2020}
A. C. Wang, F. Huang, W. L. Wang, and G. X. Peng, Phys. Rev. C {\bf 102}, 015203 (2020).
%
\bibitem{Huang:2009}
F. Huang, A. Sibirtsev, S. Krewald, C. Hanhart, J. Haidenbauer, and U.-G. Mei{\ss}ner, Eur. Phys. J. A {\bf 40}, 77 (2009).
%
\bibitem{Sibirtsev:2009}
A. Sibirtsev, J. Haidenbauer, F. Huang, S. Krewald, and U.-G. Mei{\ss}ner, Eur. Phys. J. A {\bf 40}, 65 (2009).
%
\bibitem{Huang:2010}
F. Huang, A. Sibirtsev, J. Haidenbauer, S. Krewald, and U.-G. Mei{\ss}ner, Eur. Phys. J. A {\bf 44}, 81 (2010).
%
\bibitem{Guidal:1997}
M. Guidal, J. M. Laget, and M. Vanderhaeghen, Nucl. Phys. A {\bf 627}, 645 (1997).
%
\bibitem{Toki:2008}
H. Toki, C. Garcia-Recio, and J. Nieves, Phys. Rev. D {\bf 77}, 034001 (2008).
%
\bibitem{Nam:2010}
S. I. Nam and C. W. Kao, Phys. Rev. C {\bf 81}, 055206 (2010).
%
\bibitem{He:2012}
J. He and X. R. Chen, Phys. Rev. C {\bf 86}, 035204 (2012).
%
\bibitem{Sokhoyan:2015}
V. Sokhoyan {\it et al.} (CBELSA/TAPS Collaboration), Eur. Phys. J. A {\bf 51}, 95 (2015).
%
\end{thebibliography}
\end{document}